\documentclass[twocolumn,prl,aps,showpacs]{revtex4-1}

\usepackage{graphicx}
\usepackage{color}
\usepackage[T1]{fontenc}
\usepackage{bm}

\newcommand{\beq}{\begin{eqnarray}}
\newcommand{\eeq}{\end{eqnarray}}

\begin{document}
\setcounter{page}{1}

\title{Ultradilute quantum liquid drops
}
\author{V. Cikojevi\'{c}, K. D\v{z}elalija, P. Stipanovi\'{c}, L. Vranje\v{s} 
Marki\'{c}}
\affiliation{Faculty of Science, University of Split, Ru\dj era 
Bo\v{s}kovi\'{c}a 33, HR-21000 Split, Croatia}
\author{J. Boronat}
\affiliation{Departament de F\'{\i}sica, 
Universitat Polit\`ecnica de Catalunya, 
Campus Nord B4-B5, E-08034 Barcelona, Spain}

\begin{abstract}
Using quantum Monte Carlo methods we have studied dilute Bose-Bose mixtures 
with attractive interspecies interaction in the limit of zero temperature. The 
calculations are exact within some statistical noise and thus go beyond previous 
perturbative estimations. By tuning the intensity of the attraction,  we observe 
the evolution of an $N$-particle system from a gas to a self-bound liquid drop. This 
observation agrees with recent experimental findings and allows for the study of 
an ultradilute liquid never observed before in Nature.  
\end{abstract}


\maketitle

The high tunability of interactions in ultracold Bose and Fermi gases is 
allowing for exploration of regimes and phases difficult to find in other 
condensed-matter systems~\cite{pethick}. By adjusting properly the applied 
magnetic field, 
Bose and Fermi gases are driven to Feshbach resonances with an increase of 
interaction practically at will, and with the possibility of turning the system 
from repulsive to attractive and vice versa. This is obviously not possible in 
conventional condensed matter where interactions are generally not tunable at 
this level. A significant example of this versatility has been the clean 
experimental realization of the unitary limit for fermions~\cite{bec1,bec2} and 
the precise 
characterization of the BCS-BEC crossover~\cite{carlson,astra}, which up to 
that moment, was only a theoretical scenario.

Recently, it has been possible to explore the formation of liquid/solid 
patterns in dilute gases by modifying the strength of the short-range 
interatomic interactions. Probably, the most dramatic example of this progress
has been the observation of the Rosensweig instability in a confined system of 
$^{164}$Dy atoms with a significant magnetic dipolar moment~\cite{kadau}.  By 
tuning the short-range interaction, Kadau \textit{et al.}~\cite{kadau} have observed  the 
spontaneous formation of an array of self-bound droplets remembering the 
characteristics of a classical ferrofluid. The observation of solid-like 
arrangements in dilute gases has also been possible working with highly-excited 
Rydberg atoms~\cite{schauss}. By direct imaging, Schauss \textit{et 
al.}~\cite{schauss} have obtained ordered 
excitation patterns with a geometry close to the well known arrangements 
observed in few-body confined Coulomb particles.

In the line of obtaining other \textit{dense} systems starting from extremely 
dilute Bose and Fermi gases, it is noticeable the mechanism suggested by Petrov 
relying on Bose-Bose mixtures~\cite{petrov}. According to this proposal, it is 
possible to 
stabilize a mixture with attractive interspecies interaction in such a way that 
the resulting system is self-bound, i.e., a liquid. Whereas a mean-field 
treatment of the mixture predicts a collapsed state, the first beyond mean 
field correction, the Lee-Huang-Yang (LHY) term, is able to stabilize the 
system by properly selecting the interspecies $s$-wave scattering length. 
Further work has shown that reducing the dimensionality of the setup to two 
or quasi-two dimensions may help to stabilize the liquid phase~\cite{petrov2}. 
The LHY 
correction has also been used to account for the formation of dipolar 
drops~\cite{santos} and 
then confirmed by full first-principles quantum Monte Carlo (QMC) 
simulations~\cite{saito,macia}.

The exciting idea of producing self-bound liquid drops by using interspecies 
attractive interaction acting as glue of the entire Bose-Bose mixture has been 
put forward by Tarruell and collaborators~\cite{tarruell}. Results 
obtained with a  mixture of 
$^{39}$K atoms in different hyperfine states have shown the formation of these drops, that do not release 
for a significant time when the confining trap is removed. Therefore, the 
theoretical prediction seems confirmed and thus a new window for exploring 
matter in unprecedented situations is open. On one side, it proves the way of 
forming liquid drops with large density in the world of cold gases and, on the 
other, makes possible the study of a liquid state of matter with an extremely 
low density, lower than any other existing liquid.     

In the present work, we study the formation of liquid drops in a Bose-Bose 
mixture using the diffusion Monte Carlo (DMC) method, which solves 
stochastically the $N$-body Schr\"odinger equation in an exact way within some 
statistical uncertainties. The DMC method was extensively used in the 
past for determining the structure and energy properties of liquid drops of 
$^4$He~\cite{pandha,chin}, $^3$He~\cite{guardiola,boro}, H$_2$~\cite{ceper}, 
and spin-polarized tritium~\cite{leandra}. At difference with previous 
perturbative estimates, DMC allows for an exact study of the quantum properties 
of the system relying only on its Hamiltonian. Our results confirm the LHY 
prediction on the stability of self-bound mixtures and determine quantitatively 
the conditions under which liquid drops are stable and how they evolve when the 
attractive interaction is increased. Within the regime here explored, we do not 
observe a full collapse of the drop but an increase of the density and 
reduction of the size, which is rather progressive.

The Bose-Bose mixture under study is composed of $N_1$ bosons of mass $m_1$ and 
$N_2$ bosons of mass $m_2$ with Hamiltonian
\begin{eqnarray}
H & = &  -\frac{\hbar^2}{2m_1} \sum_{i=1}^{N_1} {\bm \nabla}_i^2 
-\frac{\hbar^2}{2 m_2} \sum_{j=1}^{N_2} {\bm \nabla}_j^2 
 \nonumber \\
& & + \frac{1}{2} 
\sum_{\alpha,\beta=1}^{2} \sum_{i_\alpha,j_\beta=1}^{N_\alpha,N_\beta} 
V^{(\alpha,\beta)}(r_{i_\alpha j_\beta}) \ ,
\label{hamiltonian}
\end{eqnarray}
with $V^{(\alpha,\beta)}(r)$ the interatomic interaction between species 
$\alpha$ and $\beta$. Our interest is focused on a mixture of intraspecies 
repulsive interaction, i.e., positive $s$-wave scattering lengths $a_{11}>0$ 
and $a_{22}>0$, and interspecies attractive potential, $a_{12}<0$. To set up 
this regime, we use a hard-sphere potential of radius $a_{\alpha \alpha}$ for 
potentials $V^{(\alpha,\alpha)}(r)$ and an attractive square well of depth 
$-V_0$ and range $R$ for $V^{(\alpha,\beta)}(r)$. In the latter case, we fix 
$R$ and change $V_0$ to reproduce the desired negative scattering length; 
notice that we work with negative  $a_{\alpha \beta}$ values and thus the 
attractive potential does not support a pair bound state.

The DMC method uses a guiding wave function as importance sampling to reduce 
the variance to a manageable level. We adopt a Jastrow wave function in the 
form
\begin{equation}
\Psi(\bm{R})= \prod_{1=i<j}^{N_1} f^{(1,1)}(r_{ij}) \prod_{1=i<j}^{N_2 
}f^{(2,2)}(r_{ij}) \prod_{i,j=1}^{N_1,N_2} f^{(1,2)}(r_{ij}) \ ,
\label{trialwf}
\end{equation}
with $\bm{R}=\{\bm{r}_1,\ldots,\bm{r}_{N}\}$, $N=N_1+N_2$. In the case of equal particles 
the Jastrow factor is taken as the scattering solution, 
$f^{(\alpha,\alpha)}(r)=1-a_{\alpha \alpha}/r$ for $r \ge a_{\alpha \alpha}$ 
and zero otherwise. If the pair is composed of different particles then we take 
$f^{(1,2)}(r)= \exp(-r/r_0)$, with $r_0$ a variational parameter.

\begin{figure}[t]
\begin{center}
\includegraphics[width=0.9\linewidth,angle=0]{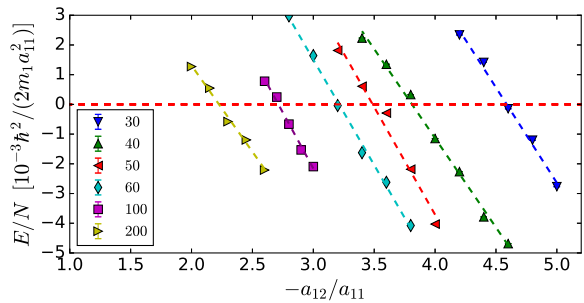}
\caption{(Color online). Energy per particle of the Bose-Bose mixture as a 
function of the scattering length $a_{12}/a_{11}$. Different symbols and lines 
correspond to DMC calculations with different number of particles.}
\label{fig1}
\end{center}
\end{figure}

In order to reduce the number of variables of the problem, keeping the 
essentials, we consider $m_1=m_2$, $N_1=N_2$, and $a_{11}=a_{22}$. In this way, 
our study explores the stability and formation of liquid drops as a function of 
$a_{12}$ and the number of particles $N$ ($N_1=N_2=N/2$). The $s$-wave 
scattering length $a_{12}$ of an attractive well is analytically 
known, $a_{12}=R \left[ 1- \tan(KR)/(KR) \right]$, with $K^2=m_1 V_0/\hbar^2$. 
We take $a_{12}<0$, which correspond to $KR < \pi/2$. In practice, we fix the 
range of the well $R$ and vary the depth $V_0$. As it is obvious from the 
equation for $a_{12}$, its value depends on the product $R V_0^{1/2}$ and then 
decreasing $R$ means to increase $V_0$. If for a fixed $a_{12}$ value we want 
to approach the limit $R \to 0$ then $V_0 \to \infty$, situation that makes our 
calculations extremely demanding in terms of accuracy and number of particles 
required to observe saturation. After some preliminary studies, we determined 
that $R=4 a_{11}$ is a good compromise  between accuracy and reliability and 
thus the major part of our results are obtained with that. In the following, unless stated otherwise, all energies and lengths are given in $\hbar^2/(2 m_1 a_{11}^2)$ and $a_{11}$ units, respectively.

\begin{figure}[t]
\begin{center}
\includegraphics[width=0.9\linewidth,angle=0]{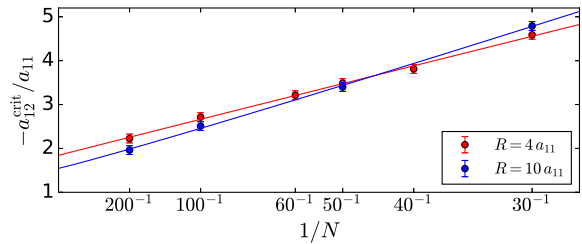}
\caption{(Color online). Critical values $a_{12}^{\rm{crit}}$ for liquid drop 
formation as a function of $1/N$. Red and blue points stand for $R=4 a_{11}$ 
and $R=10 a_{11}$, respectively. The lines correspond to linear fits to the DMC 
results.}
\label{fig2}
\end{center}
\end{figure}

The trial wave function $\Psi(\bm{R})$ (\ref{trialwf}) depends on a single 
parameter $r_0$. This parameter is previously optimized using the variational 
Monte Carlo method. Its value increases with the total number of particles $N$; 
for instance, when $R=4$ and $V_0=0.166$, $r_0$ increases monotonously from $106$ up to $622$  
when $N$ grows from $100$ to $2000$. Our DMC algorithm is accurate up to 
second order in the imaginary-time step~\cite{casu1} and uses forward walking 
to remove any 
bias of the trial wave function in the estimation of diagonal operators which 
do not commute with the Hamiltonian~\cite{casu2}. Any systematic bias derived 
from the use 
of a finite time step and a finite number of walkers in the diffusion process 
is kept smaller than the statistical noise.

In Fig. \ref{fig1}, we report results for the energy per particle of the 
Bose-Bose mixture for different number of particles and as a function of the 
scattering length $a_{12}$. For each $N$, we observe a similar behavior when we 
tune $a_{12}$. There is a critical value which separates systems with positive 
and negative energies. When the energy is positive the $N$ system is in a gas phase 
and, by increasing $|a_{12}|$, it condenses into a self-bound 
system, that is,  a liquid drop. Around the critical value the energy decreases 
linearly, the absolute value of the binding energy becoming large very quickly as a function of 
$a_{12}$. Our results show a clear dependence of the critical scattering length 
for binding on the number of particles: smaller drops require more 
attraction (larger $V_0$) than larger ones. 

\begin{figure}[t]
\begin{center}
\includegraphics[width=0.9\linewidth,angle=0]{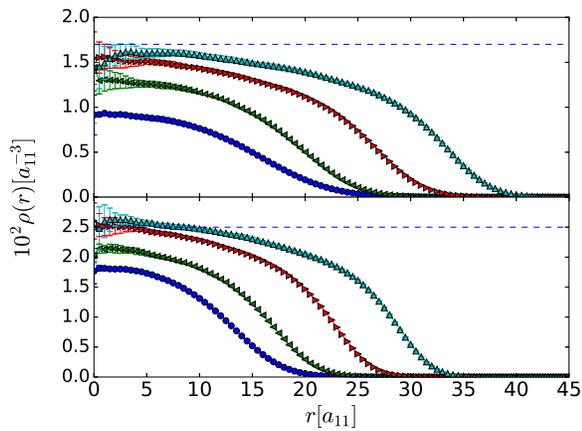}
\caption{(Color online). Density profiles of the Bose-Bose liquid drops for 
different number of particles. Top and bottom panels correspond to $V_0=0.150$, 
$a_{12}=-3.09$ and $V_0=0.166$, $a_{12}=-3.81$, respectively. From small to 
large drops, $N=200$, 400, 1000, and 2000. The dashed lines correspond to the 
equilibrium density of the bulk phase.}
\label{fig3}
\end{center}
\end{figure}

\begin{figure*}
\begin{center}
\includegraphics[width=0.9\linewidth,angle=0]{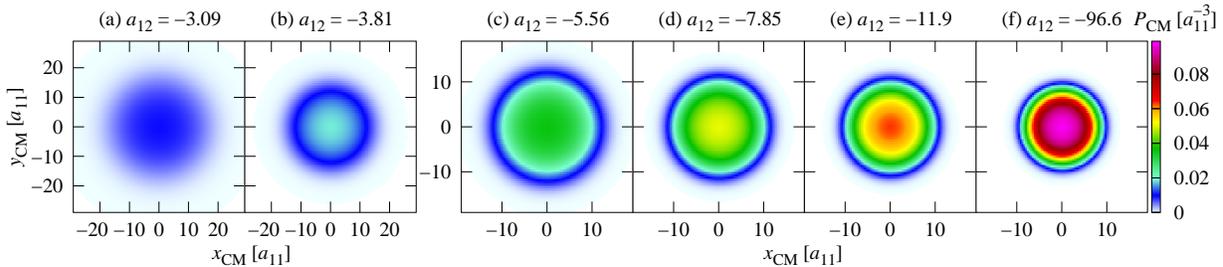}
\caption{(Color online). Contour plots of the density profiles of a liquid drop 
with $N=200$ as a function of $a_{12}$ }
\label{fig4}
\end{center}
\end{figure*}

The dependence of the critical scattering length for self binding, 
$a_{12}^{\rm{crit}}$, on the number of particles is shown in Fig. \ref{fig2}. 
Plotted as a function of $1/N$ we observe a linear decrease of  
$a_{12}^{\rm{crit}}$, reaching in the thermodynamic limit ($N \to \infty$) a 
value slightly larger than one. In fact, LHY theory has been applied to the 
formation of Bose-Bose drops around this value $|a_{12}|\sim a_{11}$ 
corresponding to drops with a very large number of particles~\cite{petrov}. In 
the same 
figure, we show results derived using a different range $R=10$ of the 
attractive well. As we can see, the results are slightly different, with an 
extrapolation to the thermodynamic limit a bit closer to one.

The calculation of the density profiles $\rho(r)$ allows for a better knowledge 
of the shape and size of the formed drops. In Fig. \ref{fig3}, we report DMC 
results on the density profiles of the obtained drops. Notice that there is no 
difference between the partial density profiles due to our election of 
interactions and masses, $\rho^{(1)}(r)=\rho^{(2)}(r)=\rho(r)/2$. The two cases 
shown in Fig. \ref{fig3} correspond to scattering lengths $a_{12}=-3.09$ (top) 
and $a_{12}=-3.81$ (bottom). When the number of particles increases one 
observes that both the central density and radius of the drop grow. This is 
expected to happen until the central density reaches the equilibrium density of 
the bulk phase. Once the drop saturates, only the radius increases with the 
addition of more particles. The density profiles, shown here for two 
illustrative examples, correspond to very dilute liquids because we need $\sim 
2000$ particles to reach saturation. In the figures, we have also shown the 
equilibrium densities that we have obtained for the same potentials in 
preliminary calculations of the bulk phase. By decreasing the scattering 
length, i.e., by making the system more attractive, we observe that the central 
density increases and the size of the drop squeezes. Apart from the central 
density one can also extract from the density profiles the surface width, 
usually measured as the length $W$ over which the density decreases from 90 to 
10\% of the inner density. It is expected that $W$ increases with $N$ for 
unsaturated drops and then it stabilizes when saturation is reached. Our 
results show also this trend: for $a_{12}=-3.09$, $W=15$ for the smallest drop 
and stabilizes then to $W\simeq 20$; for $a_{12}=-3.81$, these values are 
$W=11$ and $18$. 

DMC allows for the study of the drops around the gas-liquid transition but can also 
show how the evolution towards a collapsed state happens. By increasing the 
depth of the attractive well $V_0$ we can see the change in shape and size of a 
given drop. In Fig. \ref{fig4}, we report this evolution as a contour plot of 
the density profiles for a particular liquid drop with $N=200$ particles. The 
range of $a_{12}$ values starts close to $a_{12}^{\rm{crit}}$, for this $N$ 
value, and ends quite deep into the Feshbach resonance at a scattering length 
$a_{12} \simeq 40  \, a_{12}^{\rm{crit}}$. Following this ramp, we observe an 
increase of an order of magnitude in the inner density and a shrinking of the 
size, with a reduction of the radius in a factor of three. Therefore, the drop 
becomes more dense but it is still a fully stable object which is not 
at all collapsed.

\begin{figure}
\begin{center}
\includegraphics[width=0.9\linewidth,angle=0]{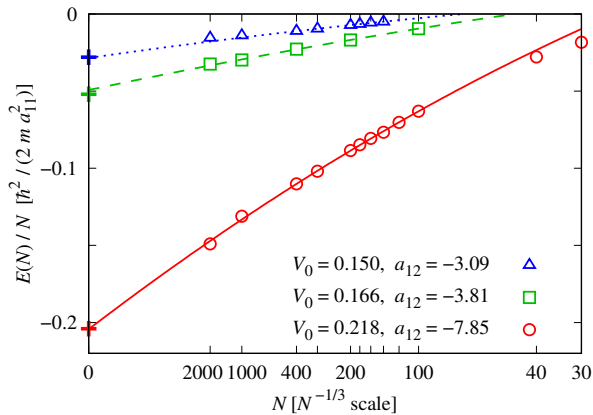}
\caption{(Color online). Energy per particle of the Bose-Bose drops as a 
function of $N^{-1/3}$. The open symbols are the DMC results and the lines are 
fits according to the liquid drop model (\ref{dropmodel}). The errorbars are 
smaller than the size of the symbols. Different sets correspond to different 
values of the interparticle scattering length $a_{12}$. The cross points at zero 
$x$-axis correspond to bulk calculations.}
\label{fig5}
\end{center}
\end{figure}

The microscopic characterization of the Bose-Bose liquid drops is not complete 
without the knowledge of the energy. As we commented before, it is the result 
of the energy which determines if an $N$-particle system is in a gas or liquid 
state. Once in the liquid phase, it is important to calculate the dependence of 
the energy on the number of particles. In Fig. \ref{fig5}, we report the DMC 
energies as a function of $N$ and for three different $a_{12}$ values. From 
intensive calculations carried out in the past on liquid $^4$He 
drops~\cite{pandha,chin}, we know 
that the energy of the drops is well accounted for by a liquid-drop model. 
According to this, the energy per particle is
\begin{equation}
E(N)/N = E_v + E_s \, x + E_c \, x^2 \ , 
\label{dropmodel}
\end{equation}
with $x\equiv N^{-1/3}$. The coefficients in Eq. (\ref{dropmodel}), $E_v$, 
$E_s$, and $E_c$ are termed volume, surface, and curvature energies, 
respectively. The term $E_v$ corresponds to the energy of an infinitely large 
drop or, in other words, to the energy per particle of the bulk. The second 
term $E_s$ is important because, from it, we can estimate the surface tension 
of the liquid $t$ as
\begin{equation} 
t = \frac{E_s}{4 \pi r_0^2} \ .
\label{tension}
\end{equation}
The parameter $r_0$ is the unit radius of the liquid, and can be estimated from 
the relation $4 \pi r_0^3 \rho_0  /3  =1$, with $\rho_0$ the equilibrium 
density of the liquid.

In Fig. \ref{fig5}, we plot as lines the results of the liquid-drop model 
obtained as least-squares fit to the DMC energies. In the three cases studied 
we obtain a high-fidelity fit. In the Figure we plot in the zero $x$-axis the 
energies of the bulk liquid in the same conditions as the drops. These 
results are not included in the fit (\ref{dropmodel}) but they are completely 
coincident with the energies $E_v$ obtained solely from the drop energies. This 
is in fact a stringent test of accuracy on the calculations of the liquid 
drops. In the figure we see the effect of the potential on the energy of the 
drops for the three selected cases. The self binding of a given $N$ drop becomes stronger as depth $V_0$, and thus $|a_{12}|$, increase. We have verified that the 
absolute value of the energy grows linearly with $V_0$ close to the critical value for self binding 
but, for larger potential depths, it increases faster. From relation 
(\ref{tension}) and the values obtained for $E_s$ from the fits using the 
liquid-drop model we estimate that the surface tension for the three cases shown 
in Fig. \ref{fig5} are $0.18 \cdot 10^{-3}$, $0.37\cdot 10^{-3}$, and $2.41 
\cdot 10^{-3}$ (in units $\hbar^2/(2 m_1 a_{11}^4)$) when $a_{12}=-3.09$, 
$-3.81$, and $-7.85$, respectively.   

We think that a comparison between the Bose-Bose drops here studied and the 
well-known properties of stable superfluid $^4$He drops can help to better 
visualize their extraordinary properties. We can consider a typical value for 
$a_{11}$ used in the experiments with ultracold mixtures of $^{39}$K, 
say $a_{11}=50 \, a_0$, with $a_0$ the Bohr radius. Then, the saturation 
densities of the drops shown in Fig. \ref{fig3} are $\sim 1.0 \cdot 10^{-6}$ and 
$ 1.4 \cdot 10^{-6}\ \text{\AA}^{-3}$. The saturation density of liquid $^4$He 
is $2.2 \cdot 10^{-2} \ \text{\AA}^{-3}$ implying that the Bose-Bose drops can 
be as dilute as $\sim 10^4$ times the $^4$He ones (a similar ratio happens when 
compared with water, with density $3.3\cdot10^{-2}\ \text{\AA}^{-3}$)~\cite{bulk}. For the 
same number of atoms, the Bose-Bose drop is much larger than the $^4$He one: 
$9.8 \cdot 10^{-2}$ $\mu$m for $V_0=0.150$ and $3 \cdot 10^{-3}$ $\mu$m for $^4$He with $N=2000$~\cite{barranco}. The surface of the dilute drop for this $N$ is $\sim 50$\% of the total size, much larger than the 20\% value in $^4$He.

Summarizing, we have carried out a DMC calculation of Bose-Bose mixtures with 
attractive interspecies interaction. Relying only on the Hamiltonian, we 
describe the system without further approximations. As announced by Petrov using 
LHY approximate theory~\cite{petrov}, it is possible to get self-bound systems 
by a proper selection of the interactions between equal and different species. 
The versatility of ultracold gases to change its interaction in magnitude 
and sign opens the possibility to explore new and very exciting physics. Our 
results clearly show the transition from a gas, with positive energy, to a 
self-bound system (liquid) and determine accurately the critical scattering 
lengths for the transition as a function of the number of particles. In the 
range of parameters here studied, we do not observe universality in the sense 
that the results depend only on the $s$-wave scattering lengths. For a same 
$a_{12}$ value we observe dependence on the range of the potential $R$. 
Preliminary calculations 
of the bulk phase seem to confirm this result, even carrying $R \to 0$. Future
studies can try to see if the inclusion of the effective range to better fit the 
model potential can help to devise a more precise description. 

The experimental realization of Bose-Bose liquid drops~\cite{tarruell} opens 
the possibility of accessing  denser systems than the usual trapped ultracold 
gases where quantum correlations can be much more relevant. The point of view 
from liquid state is however different: the liquid that emerges from these 
mixtures is ultradilute, much less dense than any other stable liquid in 
Nature. Therefore, the liquid phase realm extends to unexpected regimes never 
achieved before.

\acknowledgments
We acknowledge fruitful discussions with Leticia Tarruell and Gregory
Astrakharchik. This work has been supported in part by the Croatian Science Foundation under the project number IP-2014-09-2452.
Partial financial support from the MINECO (Spain) grant No. FIS 
2014-56257-C2-1-P is also acknowledged. 
The computational resources of the Isabella cluster at Zagreb 
University Computing Center (Srce) and the HYBRID cluster at 
the University of Split, Faculty of Science and Croatian National
Grid Infrastructure (CRO NGI) were used.


\begin{thebibliography}{99}
\bibitem{bulk} Preliminary DMC calculations of the bulk liquid show that equilibrium densities as low as $2\cdot 10^{-11} \ \text{\AA}^{-3}$ can be achieved. These would correspond to very large $N$ drops, which are hardly accesible with direct DMC simulations. These densities are even below the ones observed experimentally in Ref.\cite{tarruell}.
\end{thebibliography}

\end{document}